\documentclass[serif]{WileyMSP-template}
\usepackage[english]{babel}
\usepackage{amsmath}
\usepackage{xcolor}
\usepackage[square, sort, numbers, super,compress]{natbib}
\usepackage{upgreek}
\usepackage{graphics}
\usepackage{graphicx}
\usepackage{multicol}
\usepackage{lipsum}
\usepackage{hyperref}
\usepackage{siunitx}
\begin{document}

\pagestyle{fancy}
\rhead{\includegraphics[width=2.5cm]{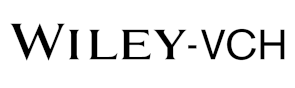}}
\title{Bidirectional optical non-reciprocity in a multi-mode cavity 
optomechanical system}

\maketitle

\author{Muhib Ullah}
\author{Xihua Yang}
\author{Li-Gang Wang*}

\begin{affiliations}
Dr. M. Ullah,\\
Department of Physics, Zhejiang University, Hangzhou 310027, China\\
Prof. X. Yang,\\
Department of Physics, Shanghai University, Shanghai  200444, China\\

Prof. L.-G. Wang\\
Department of Physics, Zhejiang University, Hangzhou 310027, China\\
Canadian Quantum Research Center, 204-3002 32 Ave, Vernon, BC V1T 
2L7, Canada\\
Email Address:lgwang@zju.edu.cn

\end{affiliations}

\begin{abstract}
\justify
Optical non-reciprocity that allows unidirectional flow of
optical field is pivoted on the time reversal symmetry breaking, which 
originates from radiation pressure because of 
light-matter 
interaction in 
cavity optomechanical systems. Here,  the non-reciprocal transport of optical 
signals across two ports
via three optical modes optomechanically coupled to the mechanical excitation of
two nanomechanical resonators (NMRs) is studied under the 
influence of strong
classical drive fields and weak probe fields. It is found that there exists the 
conversion of reciprocal to non-reciprocal signal transmission via tuning the 
drive fields and perfect
non-reciprocal transmission of output fields can be realized when the effective 
cavity
detuning parameters are near resonant to the NMRs' frequencies. The
unidirectional non-reciprocal transport is robust to the
optomechanical couplings around resonance conditions. Moreover,
the cavities' photon loss rates play an inevitable role in the
unidirectional flow of signal across the two ports. Bidirectional
transmission can also be controlled by the phase changes associated with the 
probe and drive fields along with their relative phase. 
This scheme may provide a foundation for the compact 
non-reciprocal communication and quantum information processing, thus enabling 
novel devices that route
photons in unconventional ways such as all-optical diodes, optical transistors 
and optical switches.

\end{abstract}

\section{Introduction}
\label{intro}
\justify
Non-reciprocity is a phenomenon in certain devices that allows signal to pass
through in one direction, but block it in the opposite, and is requisite in a
broad range of applications such as invisibility or cloaking, and noise free
information processing. \cite{sounas2017} Optical non-reciprocity has
originated from breaking the Lorentz reciprocity theorem.~\cite{lorentz1896}
Apart from that, optical non-reciprocity has been realized in
magneto-optical Faraday 
effect,~\cite{levy2005,zaman2007,dotsch2005,wang2009,khanikaev2010,bi2011,chin2013}
but
the major flaw in these devices is their inconvenience in integration
because of some issues such as cross talk caused by the magnetic field,
ill-suitableness for sensitive superconducting circuits as their strong
magnetic fields are highly disruptive and need strong shielding, and lattice
mismatches between magneto-optical materials and silicon.~\cite{dai2012} In
addition, magneto-optical materials manifest remarkable loss at optical
frequencies, that is, the order of 100 dB cm$^{{-}1}$, making them
sub-optimal solutions for high efficiency devices. As an 
alternate to the 
magneto-optical non-reciprocal devices, several techniques 
have been 
practiced by using microwave chip-level systems. One approach is based on an
artificial magnetic field by modulating the parametric coupling between the
modes of a network, thus making the system non-reciprocal at the 
ports.~\cite{estep2014,kerckhoff2015} Second technique is 
the phase
matching of a parametric interaction that leads to non-reciprocal behavior
of the communicating signal, since the signal only interacts with the pump
when co-propagating with it and not in the opposite direction. This causes
traveling-wave amplification to be 
directional.~\cite{ranzani2014,white2015,macklin2015,hua2016}

The approach for on-chip optical non-reciprocity has also been used recently by
using a strong optomechanical interaction between the external fields and 
micro-ring resonators,~\cite{hafezi2012} and this has been experimentally
demonstrated using a silica microsphere resonator.~\cite{shen2016} The 
optomechanical interaction basically arises from the radiation pressure between 
cavity photons and mechanical resonators in an optomechanical cavity whose 
details can be found in a recent review by Meystre.~\cite{meystre2013} Using 
the three-mode optomechanical system, Chen et al., have proposed a 
scheme for non-reciprocal mechanical squeezing due to the joint
effect of the mechanical intrinsic nonlinearity and the quadratic
optomechanical coupling.~\cite{chen2020}
In a similar fashion, an optomechanical circulator and directional amplifier
in a two-tapered fiber-coupled silica micro-resonator have been proposed to
perform as an add-drop filter, and they may be switched to
circulator mode or directional amplifier mode via a simple change in the
control field.~\cite{shen2018} It has been accredited that the
non-reciprocal signal transfer between two optical modes mediated by
mechanical mode can be realized with suitable optical driving.~\cite%
{habraken2012,xu2015}  Additionally, these modes
in cavity optomechanics can also result in some other interesting effects like 
ground-state cooling of a NMR, \cite{zhou2016,ondrej2019}
steady-state light-mechanical quantum steerable correlations in a cavity 
optomechanical system (COS), \cite{tan2017}
slow-to-fast light tuning and single-to-double optomechanically induced 
transparency (analogous to electromagnetically induced transparency), 
\cite{rahman2018} flexible manipulation on Goos-H$\ddot{\rm{a}}$nchen shift as 
a classical application
of COS, \cite{ullah2019} Fano resonances, \cite{qu2013} superradiance, 
\cite{kipf2014} optomechanically induced opacity and amplification in a 
quadratically coupled COS, \cite{si2017} and they can also be  used for 
non-classical state generation in cavity QED when atom interacts with the 
cavity dynamics to induce large nonlinearity in the system. \cite{ville2019} 
Similarly, by tailoring the fluctuations of driving fields in an optomechanical 
system with a feedback loop, the performance of optomechanical system is 
greatly improved. ~\cite{nenad2017} Apart from
that, Peterson et al., have further demonstrated an
efficient frequency-converting microwave isolator, stemmed on the
optomechanical interactions between electromagnetic fields and a
mechanically compliant vacuum-gap capacitor, which does not require a static
magnetic field and allows a dynamic control of the direction of 
isolation.~\cite{peterson2017} Bernier et al., have experimentally 
realized
the non-reciprocal scheme in an optomechanical system using a
superconducting circuit in which mechanical motion is capacitively coupled
to a multi-mode microwave circuit.~\cite{bernier2017} Similarly, Barzanjeh 
et al., have presented an on-chip microwave circulator using a 
frequency tunable silicon-on-insulator electromechanical system to investigate 
non-reciprocity via two output ports and is also compatible with 
superconducting qubits. \cite{barzanjeh2017}

Fetching an insight from the above discussion, we introduce a scheme to
achieve bidirectional non-reciprocal signal transmission using purely
optomechanical interactions in the presence of a partial beam splitter (BS).
The setup consists of two ports (left and right) through which the signal
exchange occurs. The external fields interact with the cavity modes and thus
with the nanomechanical resonators' (NMRs) phonons via 
radiation pressure
forces, which induce effective nonlinearity into the system and breaks the time
reversal symmetry. These factors are ultimately accountable for the optical
non-reciprocal behavior of the system to incoming light fields.
Non-reciprocal process as a result of interference due to different phases
has been discussed in a two-mode cavity system with two mechanical modes.~%
\cite{xu2016,tian2017}  Very recently, in a
letter, a configurable and directional electromagnetic signal transmission
has been shown to be obtained in an optomechanical system by designing a
loop of interactions in the synthetic plane generated by driven Floquet
modes on one hand and multiple mechanical modes on the other hand, to
realize a microwave isolator and a directional amplifier.~\cite{mercier2020}

This work is organized as follows. In section \ref{model}, we present the model 
of
multi-mode COS and calculate the analytical results
for the output fields of both ports 1 and 2. In section \ref{results}, we 
analyze and
discuss our results numerically and explain the non-reciprocal behavior of 
output signals
under different system parameters. In the last section \ref{summary}, we 
conclude our work.
\section{\textbf{Model and Calculations}}
\label{model}
\begin{figure}[tbhp]
	\centering
	\includegraphics[width=3.2in]{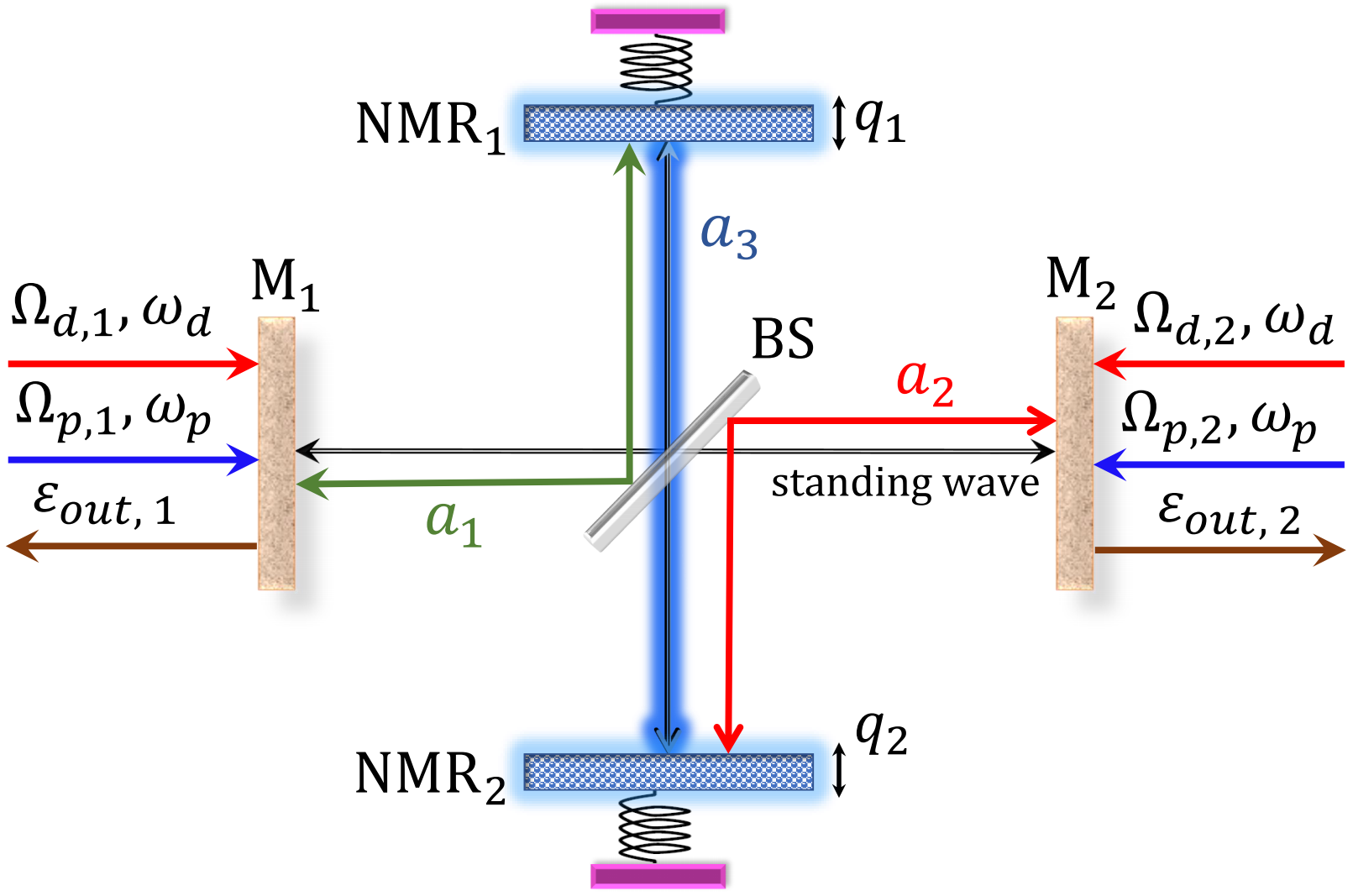}
	\caption{Schematic of a two ports multi-mode optomechanical cavity setup 
		excited by external classical fields. The setup 
		includes two fixed 
		partially transparent mirrors (M$_{1}$ and M$_{2}$) and two movable
		perfectly reflecting nano-mechanical resonators (NMR$_{1}$ and 
		NMR$_{2}$)
		with small displacements $q_1$ and $q_2$ from their respective 
		equilibrium
		positions. A partial BS is placed at the center inside the mirrors
		configuration which form three uneven cavity 
		modes. These cavity modes, that is,	$a_1$, 
		$a_2$ 
		and $a_3$, and two mechanical (phononic) modes ($b_1$ and 
		$b_2$) in this system are
		interconnected via optomechanical couplings, while a standing
		wave between M$_{1}$ and M$_{2}$ being represented 
		by a straight horizontal 
		arrow is formed.
		Four classical fields, i.e., strong drive fields with strengths 
		$\Omega_{d1}
		$ and $\Omega_{d2}$ (same frequency $\protect\omega_{d}$), and weak 
		probe fields
		having strengths $\Omega_{p1}$ and $\Omega_{p2}$ 
		(same frequency 
		$\protect\omega_{p}$) interact with the cavity system from the 
		respective sides via M$_{1}$
		and M$_{2}$, while the output fields 
		($\protect\varepsilon_{\rm{out,1}}$
		and $\protect\varepsilon_{\rm{out,2}}$) can be drawn out via left and 
		right ports, respectively.}\label{system}
\end{figure}

\justify
The proposed model shown in \textbf{Figure \ref{system}} 
is a 
two-port 
COS that is 
composed of two
partially transparent mirrors (M$_1$ and M$_2$) fixed opposite to each other
and two perfectly reflecting movable NMRs oscillating along the same axis
and a 50:50 partial BS is placed between them.
The NMRs oscillate around their equilibrium
positions with small displacements $q_1$ and $q_2$, usually in the order of 
10$^{\rm{-}9}
$ m or even \textcolor{magenta}{lower},~\cite{liu2021}  which is much smaller 
than the characteristic wavelengths of cavity modes. Thus, 
different cavity 
modes are essentially determined by their own cavity lengths.
According to Figure~\ref{system}, there are three cavity 
modes, $a_1$, $a_2$
and $a_3$, interacting with the NMRs. Modes $a_1$ and 
$a_2$ are, 
respectively, formed independently between the fixed mirrors M$_{1,2}$ and the 
NMR$_{1,2}$, while
the cavity mode $a_3$ is also formed between NMR$_{1}$ and NMR$_{2}$ via the BS.
Here we assume that all these cavity modes have different frequencies since the 
cavity lengths are unlike in general.  The last cavity 
mode between two 
fixed mirrors M$_{1}$ and M$_{2}$ can be neglected since it 
does not have any 
interaction with those NMRs.
To control or manipulate the proposed 
COS, two 
external classical and 
strong driving fields with field strengths $\Omega_{d1},~
\Omega_{d2}$ and the same frequency $\omega_{d}$ and the two weak probe fields 
with field
strengths $\Omega_{p1},~\Omega_{p2}$ and the same frequency $\omega_{p}$ are
injected from both ports (left and right) to the COS setup. Since the NMRs are 
supposed to be perfect reflecting mirrors and 
a 50:50 partial BS exists in 
the setup, all these driving and probe fields exist in the whole setup and 
interact with those three cavity modes $a_1$, $a_2$  and $a_3$. After 
interacting
with the cavity dynamics, the output probe fields ($\varepsilon_{\rm{out,1}}%
,~\varepsilon_{\rm{out,2}}$) can be collected at the left and right
ports, respectively.

The Hamiltonian for this COS in the frame rotating at the drive field
frequency $\omega_{d}$ can be given as
\begin{align}
	H_{T}=&\sum_{i=1}^{3}\Delta_{ai}a_{i}^{\dagger}a_{i}+\sum_{j=1}^{2}%
	\omega_{mj}b_{j}^{\dagger}b_{j}
	+\sum_{i=1}^{2}O_{mi}%
	a_{i}^{\dagger}a_{i}(b_{i}^{\dagger}+b_{i})
	+ O_{m31} 
	a_{3}^{\dagger}a_{3}(b_{1}^{\dagger}+b_{1})+O_{m32}a_{3}^{\dagger}a_{3} 
	(b_{2}^{\dagger}+b_{2})\nonumber \\&
	+\sum_{j=1}^{2}\sum_{k=1}^{3}i \Omega_{dj}%
	(e^{i\Phi_{dj}}a_{k}^{\dagger}  - e^{-i\Phi_{dj}}a_{k})
	+\sum_{j=1}^{2}\sum_{k=1}^{3}i {\Omega_{pj}}
	(e^{-i(\Delta_{p}t-\Phi_{pj})}a_{k}^{\dagger}-\rm{H.c.})
	\label{hamiltonian}
\end{align}
where $\Delta_{a1}=\omega_{a1}-\omega_{d}$, $\Delta_{a2}=\omega_{a2}-%
\omega_{d}$ and $\Delta_{a3}=\omega_{a3}-\omega_{d}$ are the cavity-drive
field detunings, whereas $\Delta_{p}=\omega_{p}-\omega_{d}$ denotes the
probe-drive field detuning.  In Hamiltonian of Equation 
	(\ref{hamiltonian}), the 
first term stands for the energy of cavity modes $a_{i}$ with 
$i=1,2,3$ the $i$th
cavity mode. The second term shows the energy of two
bosonic modes $b_{j}$ with $j=1,2$ 
corresponding to two NMRs. The third term accounts for the
optomechanical interactions between cavity modes ($a_{1,2}$)  and mechanical 
modes ($b_{1,2}$) that
come into existence because of radiation pressure, while the parameters $%
O_{mi}~(i=1,2)$ are the optomechanical coupling strengths between the cavity 
photons
and NMRs.
The fourth and fifth terms are associated with the optomechanical interaction
between the cavity mode $a_{3}$ and two NMRs having $O_{m31}$ and $O_{m32}$ as 
the optomechanical couplings
between them. These optomechanical couplings are crucial for the realization of 
optical non-reciprocity in our proposed cavity setup. Non-reciprocity is lost 
when these couplings vanish or become equal to zero.
Here we emphasize that the hopping interactions between different cavity modes 
via the two NMRs  cannot happen in general due to the 
uneven frequencies of 
these cavity modes~\cite{xu2016}.
The last two terms correspond to the interaction of strong classical drive
fields and weak probe fields with the cavity modes, respectively, having H.c.
as the Hermitian conjugate terms. The parameters $\Omega_{dj}$ and 
$\Omega_{pj}$ ($j=1,2$) are the drive and probe field strengths, respectively, 
and are considered to have real values for convenience.

Considering that the system may be dissipative, we use the Heisenberg's
equations of motion (so-called quantum Langevin
equation) along with damping terms given  as \cite{agarwal2010,akram2015}
\begin{equation}
	\dot{Z}=-\frac{i}{\hbar }[Z,H_{T}]-\gamma Z+N
	\label{langevin equation}
\end{equation}%
where $Z\in ~(a_{1},a_{2},a_{3},b_{1},b_{2})$ is a general operator
variable, $\gamma$ is the corresponding damping term, and the term $N$ is
the quantum white noise (Brownian noise) resulting from the interaction of 
system with external bath.
Without loss of generality, the reduced Planck's constant ($\hbar $) is
considered as equal to 1. We use Equation (\ref{langevin 
		equation}) to obtain the 
equation of motion
for every operator variable and explore the dynamics of the system. By 
substituting Equation  (\ref{hamiltonian})
into Equation  (\ref{langevin equation}) and introducing 
the factorization theorem, 
that is, $\langle ab\rangle
=\langle a\rangle \langle b\rangle $ \cite{agarwal2010,he2021}, the noise terms 
vanish since their average values lower to zero and  we obtain the following 
mean-value equations.
\begin{subequations}
	\label{MeanEq}
	\begin{align}
		\langle \dot{a}_{1}\rangle  =&-(\kappa _{1}+i\Delta _{a1})\langle
		a_{1}\rangle +iO_{m1}(\langle b_{1}^{\dagger
		}\rangle +\langle b_{1}\rangle )\langle a_{1}\rangle 
		+\sum_{j=1}^{2}{\Omega _{dj}}e^{i\Phi _{dj}}+\sum_{k=1}^{2}%
		{\Omega _{pj}}e^{-i(\Delta _{p}t-\Phi _{pk})} \\
		\langle \dot{a}_{2}\rangle  =&-(\kappa _{2}+i\Delta _{a2})\langle
		a_{2}\rangle +iO_{m2}(\langle b_{2}^{\dagger
		}\rangle +\langle b_{2}\rangle )\langle a_{2}\rangle  
		+\sum_{j=1}^{2}{\Omega _{dj}}e^{i\Phi _{dj}}+\sum_{k=1}^{2}%
		{\Omega _{pj}}e^{-i(\Delta _{p}t-\Phi _{pk})}\\
		\langle \dot{a}_{3}\rangle  =&-(\kappa _{3}+i\Delta _{a3})\langle
		a_{3}\rangle +i O_{m31} (\langle b_{1}^{\dagger }\rangle +\langle 
		b_{1}\rangle)\langle a_{3}\rangle     +i O_{m32} (\langle 
		b_{2}^{\dagger }\rangle +\langle b_{2}\rangle)\langle a_{3}\rangle  
		\sum_{j=1}^{2}{\Omega _{dj}}e^{i\Phi _{dj}} \nonumber\\& 
		+\sum_{k=1}^{2}%
		{\Omega _{pj}}e^{-i(\Delta _{p}t-\Phi _{pk})} \\
		\langle \dot{b}_{1}\rangle  =&-(\gamma _{1}+i\omega _{m1})\langle
		b_{1}\rangle +{i}O_{m31}\langle a_{3}^{\dagger
		}\rangle \langle a_{3}\rangle \\
		\langle \dot{b}_{2}\rangle  =&-(\gamma _{2}+i\omega _{m2})\langle
		b_{2}\rangle +{i}O_{m32}\langle a_{3}^{\dagger
		}\rangle \langle a_{3}\rangle
	\end{align}%
\end{subequations}
It is difficult to solve the master equation exactly because of the
existence of the nonlinear terms. Hence, we apply the linearization approach
by assuming that each operator in the system can be written as the sum of
its mean value and a small fluctuation, i.e., applying an ansatz of the form
given by
~\cite{vitali2007,huang2009,zou2011} $Z =Z_{s}+\delta Z,(Z\in 
a_{1},a_{2},a_{3},b_{1},b_{2})$,
where $Z_{s}$ stands for the steady-state value and $\delta {Z}$ for the
small fluctuations around the steady-state values of all the operator
variables under observation. The fluctuations for each variable can be
addressed as
\begin{align}
	\delta a_{1}&\rightarrow \delta \tilde{a}_{1}e^{-i\Delta _{1}t}~~\delta
	a_{2}\rightarrow \delta \tilde{a}_{2}e^{-i\Delta _{2}t}  \notag \\
	\delta a_{3}&\rightarrow \delta \tilde{a}_{3}e^{-i\Delta _{3}t}~~\delta
	b_{1}\rightarrow \delta \tilde{b}_{1}e^{-i\omega _{m1}t}  \notag \\
	\delta b_{2}&\rightarrow \delta \tilde{b}_{2}e^{-i\omega _{m2}t}
	\label{ansatz1}
\end{align}%
where $\Delta _{i}~(i=1,2,3)$ is the effective cavity detuning and $\omega
_{mj}~(j=1,2)$ is the NMR's resonance frequency. As the drive fields are
much stronger than the probe fields, we can use the conditions $\left\vert
a_{is}\right\vert \gg \delta a_{i}$ $(i=1,2,3)$ and $\left\vert
b_{js}\right\vert \gg \delta b_{j}$ ($j=1,2$) in the absence of the probe
fields $\Omega _{p1}$ and $\Omega _{p2}$, and finally get the steady-state
solutions according to the method in Ref.~\cite{agarwal2014}
\begin{subequations}
	\label{steady state}
	\begin{align}
		a_{1s}=&\frac{\Omega _{d1}e^{i\Phi _{d1}}+\Omega _{d2}e^{i\Phi 
				_{d2}}}{(\kappa _{1}+i\Delta _{1})},~b_{1s}=\frac{i O_{m31} 
				\left\vert
			a_{3s}\right\vert ^{2}}{(\gamma _{1}+i\omega _{m1})}, \\
		a_{2s}=&\frac{\Omega _{d1}e^{i\Phi _{d1}}+\Omega _{d2}e^{i\Phi 
				_{d2}}}{(\kappa _{2}+i\Delta _{2})},~b_{2s}=\frac{i O_{m32} 
				\left\vert
			a_{3s}\right\vert ^{2}}{(\gamma _{2}+i\omega _{m2})}, \\
		a_{3s}=&\frac{\Omega _{d1}e^{i\Phi _{d1}}+\Omega _{d2}e^{i\Phi 
				_{d2}}}{(\kappa _{3}+i\Delta _{3})}
	\end{align}%
\end{subequations}
were $\Delta _{1}=\Delta _{a1}-{O_{m1}}%
(b_{1s}+b_{1s}^{\ast })$, $\Delta _{2}=\Delta _{a2}-O_{m2}(b_{2s}+b_{2s}^{\ast 
})$ and $\Delta _{3}=\Delta _{a3}-{O_{m31}}(b_{1s}+b_{1s}^{\ast 
})-{O_{m32}}(b_{2s}+b_{2s}^{\ast })$ are the effective cavity detunings of the 
cavity modes $a_{1}$, $a_{2}
$ and $a_{3}$, respectively. In Equation (\ref{steady 
		state}a)-(\ref{steady 
		state}c), the expressions 
$a_{is}$ ($i=1,2,3$) and $b_{js}$ ($j=1,2$) are the steady-state solutions of 
optical modes and
mechanical modes, respectively. To find out the role of the weak probe
fields in the system dynamics, the small fluctuations are taken into
consideration by using the assumption given in Equation  
	(\ref{ansatz1}), and only slowly
moving linear terms are entertained, while fast 
oscillating terms are
ignored. Thus, the linearized equations of motion for the 
fluctuation part of the
variable operators can be derived as
\begin{align}
	\delta \dot{\tilde{a}}_{1} =&-(\kappa _{1}+i\Delta _{a1})\delta \tilde{a}%
	_{1}+iO_{m1}a_{1s}\delta \tilde{b}_{1} +\Omega _{p1}e^{i\Phi 
		_{p1}}e^{-ix_{1}t} +\Omega
	_{p2}e^{i\Phi _{p2}}e^{-ix_{1}t}
	\label{fluctuation-a1}
\end{align}%
where $x_{1}=\omega _{p}-\omega _{d}-\omega _{m1}=\Delta _{p}-\omega _{m1}$
is the probe detuning and the movable mirror resonance frequency's
difference. Expressions for $\delta \dot{\tilde{a}}_{2}$ and $\delta \dot{%
	\tilde{a}}_{3}$ can be solved alike as $\delta \dot{\tilde{a}}_{1}$, and
they are given as
\begin{align}
	\delta \dot{\tilde{a}}_{2} =&-(\kappa _{2}+i\Delta _{a2})\delta \tilde{a}%
	_{2}+iO_{m2}a_{2s}\delta \tilde{b}_{2} +\Omega _{p1}e^{i\Phi 
		_{p1}}e^{-ix_{2}t}  +\Omega
	_{p2}e^{i\Phi _{p2}}e^{-ix_{2}t} \\
	\delta \dot{\tilde{a}}_{3} =&-(\kappa _{3}+i\Delta _{a3})\delta \tilde{a}%
	_{3}+ i O_{m31} a_{3s} \delta \tilde{b}_{1}+i O_{m32} a_{3s} \delta 
	\tilde{b}_{2}+\Omega _{p1}e^{i\Phi _{p1}}e^{-ix_{3}t}+ \Omega
	_{p2}e^{i\Phi _{p2}}e^{-ix_{3}t}
\end{align}%
where the parameters $x_{2}=\Delta _{p}-\omega _{m2}$ and $x_{3}=\Delta
_{p}-\Delta _{3}$. Without loss of generality, all the cavity modes are
supposed to be driven in the mechanical red sidebands with $\Delta
_{1}=\Delta _{2}=\Delta _{3}=\omega _{m1}=\omega _{m2}=\omega _{m}$.
Therefore, $x_{1}=x_{2}=x_{3}=x$ and the system is operated in the resolved
sideband regime with the condition that $\omega _{m}\gg {\kappa _{j}}$, where $%
j=1,2,3$. With the above assumptions, the coefficients of 
mechanical
mode fluctuation operators $\delta \dot{\tilde{b}}_{1}$ and $\delta \dot{%
	\tilde{b}}_{2}$ can be simplified as
\begin{align}
	\delta \dot{\tilde{b}}_{1} &=-(\gamma _{1}+i\omega _{m1})\delta \tilde{b}%
	_{1}+{i}O_{m31}a_{3s}^{\ast }\delta \tilde{a}%
	_{3}\label{fluctuation-b1}
	\\
	\delta \dot{\tilde{b}}_{2} &=-(\gamma _{2}+i\omega _{m2})\delta \tilde{b}%
	_{2} +{i}O_{m32}a_{3s}^{\ast }\delta \tilde{a}%
	_{3}\label{fluctuation-b2}
\end{align}%
The fluctuation values of the operator variables can be further expanded to
obtain the solution easily by using the ansatz given below.%
\begin{equation}
	\label{ansatz2}
	\delta \tilde{y}=\delta \tilde{y}_{+}e^{-ixt}+\delta \tilde{y}_{-}e^{ixt}
\end{equation}%
where $\delta \tilde{y}=\delta \tilde{a}_{1},\delta \tilde{a}_{2},\delta
\tilde{a}_{3},\delta \tilde{b}_{1},\text{and}~\delta \tilde{b}_{2}$ are the 
fluctuation
variables under study. By substituting Equation  
	\eqref{ansatz2} into Equations 
	\eqref{fluctuation-a1}-\eqref{fluctuation-b2}, we
achieve the simplified fluctuation operator coefficients for the optical cavity 
modes as
\begin{align}
	\label{fluc-a1}
	\delta \tilde{a}_{1+}& =\frac{iO_{m1}a_{1s}\delta \tilde{b}%
		_{1+}+\sum\limits_{j=1}^{2}\Omega _{pj}e^{i\Phi _{pj}}}{(\kappa
		_{1}+i\Delta _{a1}-ix)}\\
	\delta \tilde{a}_{2+}& =\frac{iO_{m2}a_{2s}\delta \tilde{b}%
		_{2+}+\sum\limits_{j=1}^{2}\Omega _{pj}e^{i\Phi _{pj}}}{(\kappa
		_{2}+i\Delta _{a2}-ix)}\label{fluc-a2}\\
	\delta \tilde{a}_{3+} &=\frac{i O_{m31} a_{3s} \delta \tilde b_{1+} + i 
		O_{m32} a_{3s} \delta \tilde b_{2+}+\sum\limits_{j=1}^{2}\Omega
		_{pj}e^{i\Phi _{pj}}}{(\kappa _{3}+i\Delta _{a3}-ix)}\label{fluc-a3}
\end{align}
whereas, expressions for the coefficients associated with 
the mechanical 
mode fluctuation operators can be calculated and simplified in similar fashion 
by substituting Equation  
	\eqref{ansatz2} into 
Equation \eqref{fluctuation-a1}-\eqref{fluctuation-b2} and 
can be written as
\begin{align}
	\label{fluc-b1}
	\delta \tilde{b}_{1+}& =\frac{iO_{m31}a_{3s}^{\ast }\delta 
		\tilde{a}_{3+}}{(\gamma _{1}+i\omega _{m1}-ix)}
	\\
	\delta \tilde{b}_{2+}& =\frac{iO_{m32}a_{3s}^{\ast }\delta 
		\tilde{a}_{3+}}{(\gamma _{2}+i\omega _{m2}-ix)}\label{fluc-b2}
\end{align}

As the transmission happens via fixed mirrors (left $\text{M}_{1}$ and right
$\text{M}_{2}$) that are connected to the cavity modes $a_{1}$ and $a_{2}$,
respectively, we calculate the corresponding coefficients $\delta \tilde{a}%
_{1+}$ and $\delta \tilde{a}_{2+}$. Therefore, we apply a lengthy and
tiresome but straightforward substitution method to 
Equations (\ref{fluc-a1})-(\ref{fluc-b2}) and
obtain the required analytical expressions for $\delta \tilde{a}%
_{1+}$ and $\delta \tilde{a}_{2+}$ as
\begin{align}\label{delta-a1}
	\delta \tilde{a}_{1+}=-\frac{D[ V_{2} \zeta+\left|a_{3s}\right|^2(O_{m31}^2 
		V_{2}+O_{m32}^2 V_{1}) ]}{ U_{1}[U_{3}V_{1}V_{2}+ \left|a_{3s}\right|^2 
		(O_{m31}^2 V_{2}+O_{m32}^2 V_{1})] }
\end{align}
\begin{align}\label{delta-a2}
	\delta \tilde{a}_{2+}=-\frac{D[ V_{1} 
		\zeta^{\prime}+\left|a_{3s}\right|^2(O_{m31}^2 V_{2}+O_{m32}^2 V_{1}) 
		]}{ 
		U_{2}[U_{3}V_{1}V_{2}+ \left|a_{3s}\right|^2 (O_{m31}^2 V_{2}+O_{m32}^2 
		V_{1})]}
\end{align}
where $\zeta=U_{3}V_{1}-O_{m1}O_{m31}a_{1s}a_{3s}^{*}$, 
$\zeta^{\prime}=O_{m2}O_{m32}a_{2s}a_{3s}^{*}+U_{3}V_{2}$,  
$D=\Omega_{p1}e^{i\Phi_{p1}}+\Omega_{p2}e^{i\Phi_{p2}}$,
while $U_{1}=ix- i \Delta_{a1}- \kappa_{1}$, $U_{2}=ix- i \Delta_{a2}- 
\kappa_{2}$, $U_{3}=i x- i \Delta_{a3}- \kappa_{3}$, $V_{1}=i x- i \omega_{m1}- 
\gamma_{1}$ and $V_{2}=i x- i \omega_{m2}- \gamma_{2}$ are the parametric 
symbols used in Equation (\ref{delta-a1}) and  
	(\ref{delta-a2}).

To obtain output fields ($E_{\text{out},1}$ and $E_{\text{out%
	},2}$) and study its non-reciprocal behavior through both the output ports 
in such
an optomechanical system, input-output relation is convenient to be used as
follows. \cite{walls1994,yan2014,agarwal2014}
\begin{equation}\label{general in-out}
	E_{\text{out}}(t)+E_{\text{in}}(t)=2\kappa_j \langle \delta 
	\tilde{a}_{j}\rangle
\end{equation}%
where $(j=1,2)$ and expression 
$E_{\text{out}}(t)=E_{\text{out+}}e^{-ixt}+E_{\text{out}-}e^{ixt}$ is
the output field, generally speaking, and $E_{\text{in}}=\Omega _{pj}e^{-ixt}
$ ($j=1,2$) is the input probe light field signal expression entering the 
system from both ports, while $2\kappa_j
\langle {\delta \tilde{a}_{j}}\rangle$ are the output field coefficients at 
their respective ports. By putting
the values of above parameters in Equation  (\ref{general 
		in-out}), we obtain the 
explicit
input-output relation for the system under study as
\begin{equation}\label{input-output relation}
	E_{\text{out}j+}e^{-ixt}+E_{\text{out}j-}e^{ixt}+\Omega
	_{pj}e^{-ixt}=2\kappa _{j}\langle \delta \tilde{a}_{j}\rangle 
\end{equation}%
where $(j=1,2)$ and $\langle \delta \tilde{a}_{j}\rangle =\delta \tilde{a}%
_{j+}e^{-ixt}+\delta \tilde{a}_{j-}e^{ixt}$. By replacing the value of $%
\langle \delta \tilde{a}_{j}\rangle $ in Equation  
	(\ref{input-output relation}), we 
obtain the output field expressions
for both routes, i.e., ports 1 and 2
\begin{eqnarray}
	\label{port1 output}
	E_{\text{out}j+}e^{-ixt}+E_{\text{out}j-}e^{ixt}&+\Omega _{pj}e^{-ixt}
	=2\kappa _{j}\delta \tilde{a}_{j+}e^{-ixt} +\delta 
	\tilde{a}_{j-}e^{ixt}
\end{eqnarray}%
Equating both sides of Equation  (\ref{port1 output}) with 
respect to $e^{-ixt}$ we 
obtain the output field
expression at port 1 as
\begin{equation}\label{port1 final}
	E_{\text{out}1+}=\varepsilon _{\text{out},1}=2\kappa _{1}\delta \tilde{a}%
	_{1+}-\Omega _{p1}
\end{equation}%
Similarly, for port 2 the output field relation can be derived as
\begin{equation}\label{port2 final}
	E_{\text{out}2+}=\varepsilon _{\text{out},2}=2\kappa _{2}\delta \tilde{a}%
	_{2+}-\Omega _{p2}
\end{equation}%

The expressions of the transmission amplitudes of both ports are given as 
\cite{xu2018}
\begin{eqnarray}\label{transmission 1 to 2}
	T_{2\rightarrow 1} &=&\left\vert \varepsilon _{\text{out},1}/\Omega
	_{p2}\right\vert ^{2}=\left\vert \frac{2\kappa _{1}\delta \tilde{a}%
		_{1+}-\Omega _{p1}}{\Omega _{p2}}\right\vert ^{2} \\
	T_{1\rightarrow 2} &=&\left\vert \varepsilon _{\text{out},2}/\Omega
	_{p1}\right\vert ^{2}=\left\vert \frac{2\kappa _{2}\delta \tilde{a}%
		_{2+}-\Omega _{p2}}{\Omega _{p1}}\right\vert ^{2}\label{transmission 2 
		to 1}
\end{eqnarray}%
where the strengths of probe light field injected to the system from either
port are considered same, quantitatively.
\section{\textbf{Results and Discussion}}
\label{results}
\begin{figure*}[tph]
	\centering
	\includegraphics[width=0.65\linewidth]{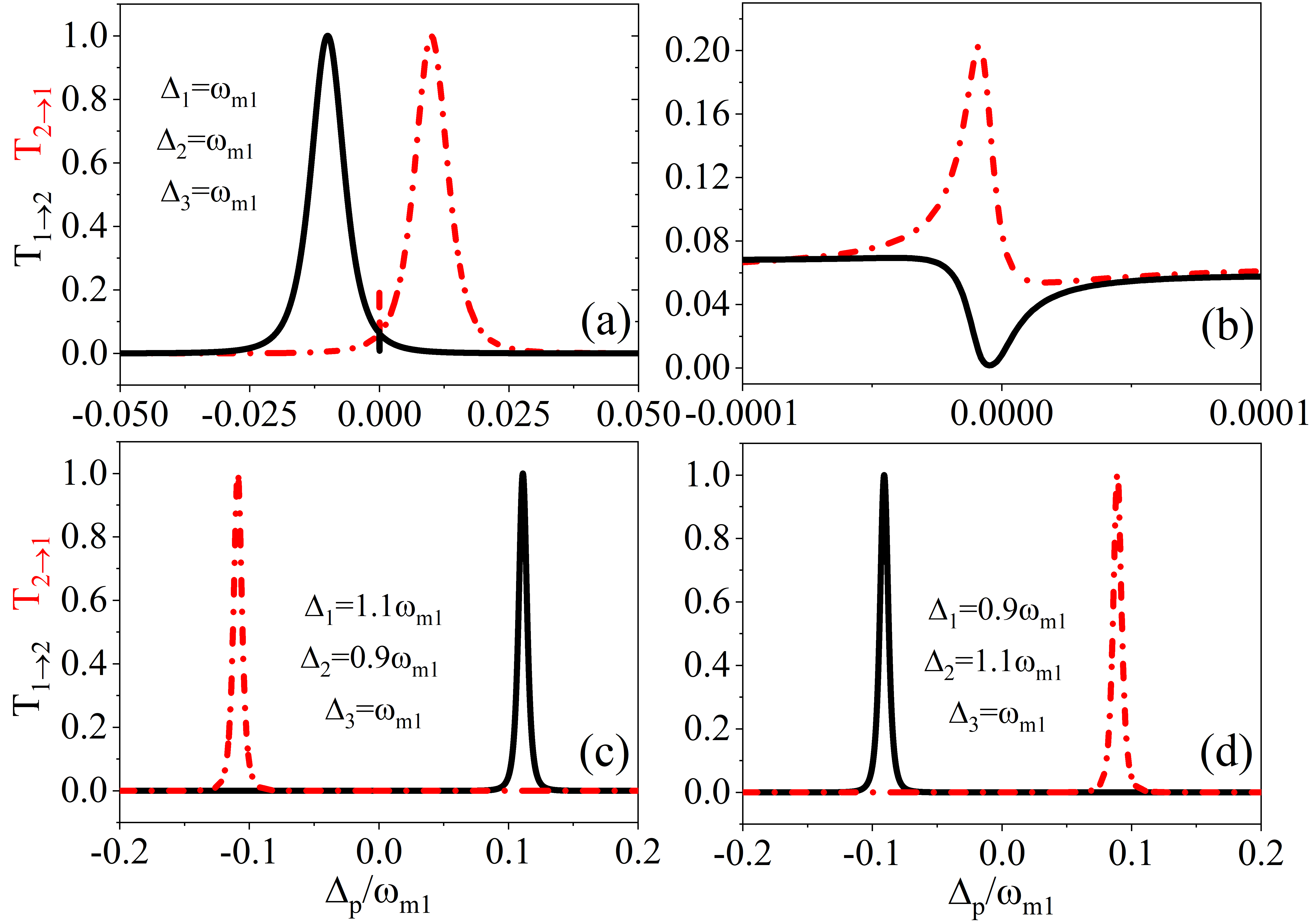}
	\caption{Transmission intensities T$_{2\rightarrow 1}$ (red dash-dot 
		curves) and T$_{1 \rightarrow 2}$ (black solid curves) as a function of 
		the 
		probe-drive field detuning $\Delta_{p}$ under different values of the 
		effective cavity detunings: (a)  the same values of effective cavity 
		detunings $\Delta_1=\Delta_2=\Delta_3= \omega_{m1}$, (b) the inset for 
		a 
		short frequency range showing the smaller dip and peak of intensity 
		profile 
		near the origin of (a), (c) $\Delta_1=1.1 \omega_{m1}$, $\Delta_2=0.9 
		\omega_{m1}$, $\Delta_3= \omega_{m1}$, and (d) $\Delta_1=0.9 
		\omega_{m1}$, 
		$\Delta_2=1.1 \omega_{m1}$, $\Delta_3= \omega_{m1}$. The general 
		parameters 
		are given as $\protect%
		\omega_{m1}/2\protect\pi=\protect\omega_{m2}/2\protect\pi= 12.6$ GHz, 
		$\protect%
		\kappa_{1}/2\protect\pi=\protect\kappa_{2}/2\protect\pi=\protect\kappa_{3}/2%
		\protect\pi=73$ MHz, 
		$\protect\gamma_{1}/2\protect\pi=\protect\gamma_{2}/2%
		\protect\pi=88$ kHz, $\Delta_{a1}/2\pi=79.96$ GHz, 
		$\Delta_{a2}/2\pi=78.38$ GHz, $\Delta_{a3}/2\pi=84.71$ GHz, 
		$O_{m1}/2\protect\pi=O_{m2}/2\protect\pi=O_{m31}/2%
		\protect\pi= O_{m32}/2%
		\protect\pi=1.5$ MHz, $\textit{L}_{i}=\textit{L}_{3i}=5.19$ mm 
		($i=1,2$), 
		$m_{\text{eff},j}=20~\upmu$g ($j=1,2$),
		$\Phi_{d1}=\Phi_{d2}=\Phi_{p1}=\Phi_{p2}=0$, $\Omega_{d1}=\Omega_{d2}=2
		\protect\omega_{m1}$, and $\Omega_{p1}=\Omega_{p2}=0.2
		\protect\omega_{m1}$.}
	\label{effective detuning}
\end{figure*}
In this section, we will numerically investigate the non-reciprocal behavior
of the output signals using the COS scheme with two signal exchange ports.
The vital role responsible for this phenomenon is played by the
optomechanical interactions between the cavity photons and 
their respective
NMRs' phonons. It is worth noting that we have considered different (unequal) 
values for the three cavity detunings ($\Delta_{a1},~\Delta_{a2}$ and 
$\Delta_{a3}$) which disregards or eliminates the possibility of photon hopping 
from one into another cavity. For numerical
simulations, we consider the practically realizable parameters from a recent 
experimental work whose values are given
as \cite{kharel2019} $\protect%
\omega_{m1}/2\protect\pi=\protect\omega_{m2}/2\protect\pi= 12.6$ GHz, $\protect%
\kappa_{1}/2\protect\pi=\protect\kappa_{2}/2\protect\pi=\protect\kappa_{3}/2%
\protect\pi=73$ MHz, $\protect\gamma_{1}/2\protect\pi=\protect\gamma_{2}/2%
\protect\pi=88$ kHz, $O_{m1}/2\protect\pi=O_{m2}/2\protect\pi=O_{m31}/2%
\protect\pi= O_{m32}/2%
\protect\pi=1.5$ MHz, $\Delta_{a1}/2\pi=79.96$ GHz, $\Delta_{a2}/2\pi=78.38$ 
GHz, $\Delta_{a3}/2\pi=84.71$ GHz,  $L_{i}=L_{3i}=5.19$ mm ($i=1,2,$), 
$m_{\text{eff},j}=20~\upmu$g ($j=1,2$),
$\Phi_{d1}=\Phi_{d2}=\Phi_{p1}=\Phi_{p2}=0$, $\Omega_{d1}=\Omega_{d2}=2
\protect\omega_{m1}$, and $\Omega_{p1}=\Omega_{p2}=0.2
\protect\omega_{m1}$. Our proposed COS can offer an excellent control and
manipulation ability to the non-reciprocal transmission. This proposal could be 
very
critical in the quantum information processing, optical sensors, optical
switches, isolators, full-duplex signal transmission and upcoming quantum 
nanotechnologies.
\subsection{Tuning $\Delta_{1}$ and $\Delta_{2}$ to control non-reciprocity}
The non-reciprocal phenomenon discussed here is based on the interference
effect at near resonance conditions. The effective cavity detunings 
$\Delta_{i}$ (
$i=1,2$) play a basic role in controlling the signal transmission. A slight
change in the values of $\Delta_{i}$ from the resonance value brings in a
perfect non-reciprocal transmission around the origin as shown in 
\textbf{Figure 
	\ref{effective detuning}}. First, we choose the values of effective cavity 
detunings to be at exact resonance, that is, 
$\Delta_{1}=\Delta_{2}=\Delta_{3}=\omega_{m1}$. Figure \ref{effective 
	detuning}(a) reveals the corresponding result showing the non-reciprocal 
behavior of signal at ports 1 and 2 plotted by red and black curves, 
respectively. Both the curves are 
close to each other on the frequency axis 
separated by a small spike-like pattern that shows non-reciprocity of signal 
curves at their extremes. The spike-like curve is enlarged to have a clear 
picture for better understanding as shown in Figure 
	\ref{effective detuning}b. 
The non-reciprocal behavior of signal inside the cavity setup happens owing 
to the quantum interference phenomenon between the fields in the optomechanical 
system. Now, by
choosing the values $\Delta_{1}=1.1 \omega_{m1}$ and $\Delta_{2}=0.9%
\omega_{m1}$, a perfect blockade of the probe signal T$_{1\rightarrow2}$ and
transmission T$_{2\rightarrow1}$ close to $\Delta_{p}=\text{-}0.1\omega_{m1}$ 
(where the peak lies) on
the frequency axis is achieved as depicted by the red curve shown in 
Figure 
	\ref{effective detuning}c.
Likewise, near $%
\Delta_{p}=0.1\omega_{m1}$ on the positive frequency axis away from the origin, 
scenario changes and
the signal transfer T$_{1\rightarrow2}$ is permitted while T$_{2\rightarrow1}
$ is completely blocked. To fully uncover the contribution of $\Delta_{i}$ to
the non-reciprocity phenomenon, the values are chosen to be $\Delta_{1}=0.9
\omega_{m1}$ and $\Delta_{2}=1.1\omega_{m1}$, so the transmission curve 
positions for
both T$_{1\rightarrow2}$ and T$_{2\rightarrow1}$ on the frequency axis are 
switched oppositely to
the previous case [see Figure \ref{effective detuning}c] 
and shift towards the 
origin as shown in Figure \ref{effective detuning}d. In 
both sub
Figures mentioned above, the probe-field transfer via 
either port occurs because 
of
constructive interference between the probe field-induced cavity field and the 
NMRs' excitations (resonance frequencies), while the transmission blockade 
comes into play due to the
destructive interference happening at the near-resonant conditions, and thus no 
probe signal is received at the output port. There is no signal transfer seen 
at either port for the frequencies other than mentioned above.  Moreover, these 
interference patterns depend on the cavity detunings, since the radiation 
pressure varies with the change in $\Delta_i$ value which 
ultimately is 
accountable for breaking the time reversal symmetry and we obtain the 
non-reciprocal transmission. Hence, by tuning the effective cavity detunings as 
near-resonant with the NMRs' excitations, the non-reciprocal output signal 
transfer via output ports can observed at a certain frequency range by using 
our proposed setup. The above discussion manifests non-reciprocity when the 
effective cavity detunings are slightly off-resonant with the mechanical 
excitations, and at exact resonance case ($\Delta_{i}=\omega_{m1}$) the signal 
non-reciprocal behavior is enhanced on account of increasing linewidth of the 
transmission curves.
Hence the effective cavity detuning can be used to flexibly control the
bidirectional output-signal transfer at either port as demanded.
\subsection{Influence of optomechanical couplings on signal transmission}
\begin{figure*}[tbhp]
	\centering
	\includegraphics[width=0.65\linewidth]{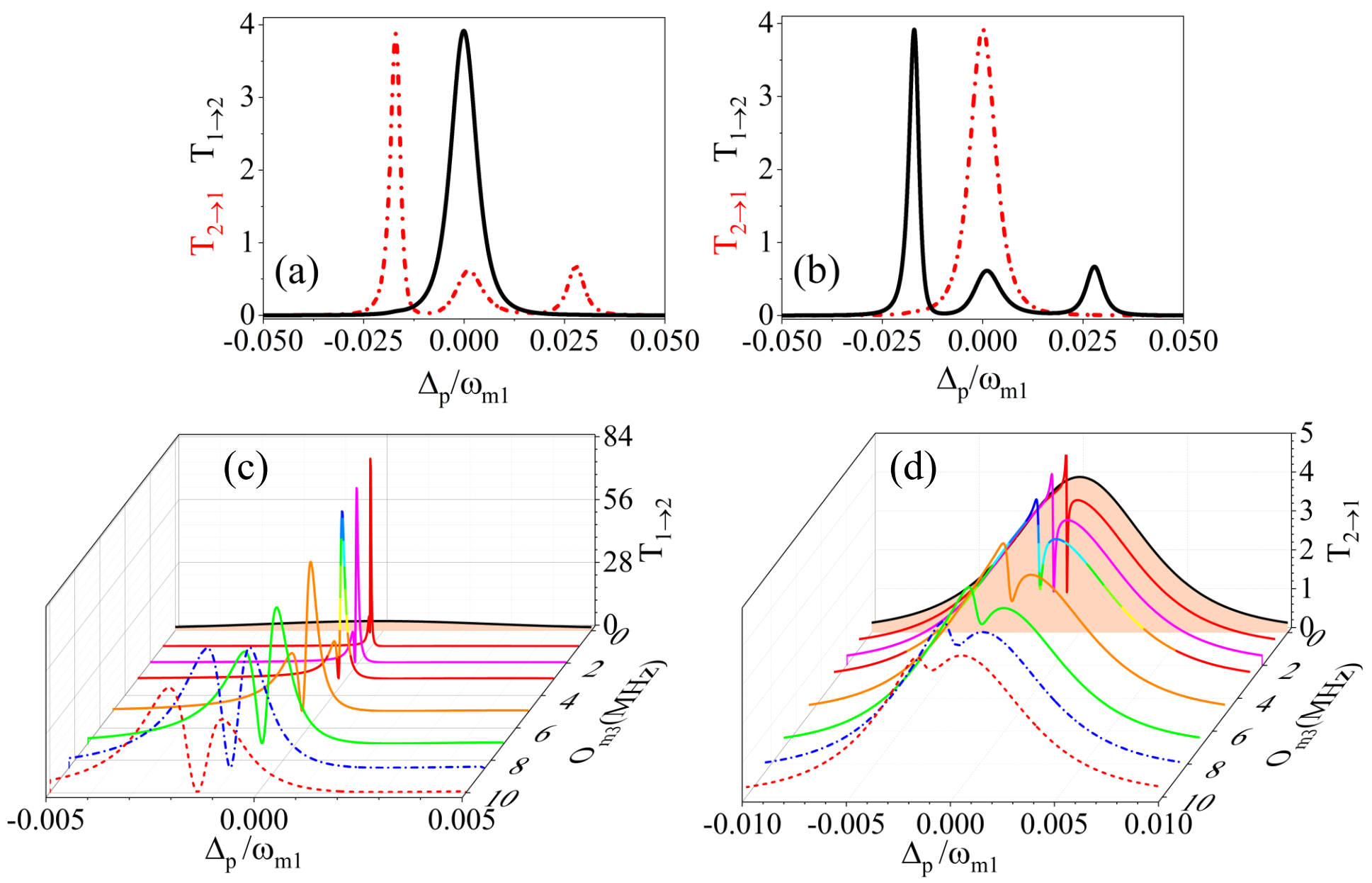}
	\caption{(Color online) (a) (b) Transmission intensities  T$_{2\rightarrow 
			1}$ (red dash-dot curves) and T$_{1 \rightarrow 2}$ (black solid 
			curves)  
		as a function of the probe-drive field detuning $\Delta_{p}$ under 
		different values of optomechanical coupling strengths $O_{m1}$, $%
		O_{m2}$, $O_{m31}$ and $O_{m32}$: (a) $O_{m1}/2\protect\pi=1$ MHz, 
		$O_{m2}/2\protect\pi=60$ MHz, $O_{m31}/2%
		\protect\pi=O_{m32}/2\protect\pi= O_{m3}/2\protect\pi= 48.5$ MHz, (b) 
		$O_{m1}/2\protect\pi=60$ MHz, $O_{m2}/2\protect\pi=1$ MHz, $O_{m31}/2%
		\protect\pi=O_{m32}/2\protect\pi=O_{m3}/2\protect\pi= 48.5$ MHz. (c)(d) 
		The waterfall plots of transmission intensities T$_{1\rightarrow 2}$ 
		and T$_{2\rightarrow 1}$ as a function of probe-drive field detuning 
		$\Delta_{p}$ and $O_{m3}$ . The optomechanical couplings for (c), (d) 
		are $O_{m1}/2\protect\pi=1$ MHz, $O_{m2}/2\protect\pi=60$ MHz. The 
		general parameters are given as  $\Delta_{a1}/2\pi=79.168$ GHz, 
		$\Delta_{a2}/2\pi=79.160$ GHz, $\Delta_{a3}/2\pi=79.96$ GHz. Other 
		values are same as mentioned in Figure 2a.}
	\label{optomechanical coupling}
\end{figure*}
In cavity optomechanics, the optomechanical coupling strength between the
intra-cavity photons and the NMR (which results from the radiation pressure
of cavity photons on the NMR) plays a key role in inducing the nonlinearity
into the cavity optomechanical system. Here we study the influence of
optomechanical coupling strength in inducing and controlling the
bidirectional non-reciprocal response of the output probe fields across two
available routes (ports). Unlike the Faraday's effect in magneto-optical
materials that makes the time-reversal symmetry breaking happen, \cite%
{chin2013} the non-reciprocity in our proposed system arises due to the
asymmetric radiation pressure of cavity photons on the NMRs when the
optomechanical couplings are unequal. The interaction between cavity fields and 
mechanical oscillations forms a five-mode chain, i.e., $a_1 \leftrightarrow b_1 
\leftrightarrow a_3 \leftrightarrow b_2 \leftrightarrow a_2$, thus operating in 
a closed loop pattern that posses quantum interference between the optical 
modes and mechanical excitations, and hence relying on constructive/destructive 
interference the flow/blockade of signal transfer becomes possible. The cavity 
mode $a_3$  in the middle (between NMRs) maintains a connection between the two 
mechanical modes and its optomechanical coupling strength plays crucial role 
regarding the behavior of signal at output ports. When the value of 
optomechanical
interaction between cavity $a_1$  and NMR$_{1}$ is $O_{m1}=1$  MHz and 
relatively higher for cavity $a_2$ , that is, $O_{m2}=60$  MHz, it
exhibits the optical signal transfer at port 1 without any
restriction around the origin (shown by red dot-dashed peak), whereas in the 
same frequency range, the transmission of signal T$%
_{1\rightarrow2}$ at port 2 is almost blocked as shown by the black
solid curve depicted in \textbf{Figure \ref{optomechanical 
coupling}a}.The physical 
picture for non-reciprocity comes from quantum interference among optical and 
mechanical modes in a similar fashion to the loop coupling of fields explained 
in Ref. \cite{xu2016}.  The non-reciprocity comes into play when the radiation 
pressure on the mechanical resonators is not even. In that case, the 
constructive interference among the mechanical excitations and optical driven 
fields leads to a high efficiency  amplified output signal, whereas destructive 
interference brings the output down to zero. In the case, the radiation 
pressure remains  same for all optomechanical couplings, no breaking of time 
reversal symmetry happens and thus signal behavior is completely reciprocal. 
Due to destructive interference, i.e., in case of $O_{m1} <
O_{m2}$, the signal is blocked in the T$_{1\rightarrow2}$ direction, whilst
at the same time, the signal transit becomes viable due to the reverse
effect, that is, constructive interference around the resonance point. The 
constructive/destructive
interference effect between the mechanical excitations and probe-induced cavity 
field happens here due to the asymmetry of the radiation pressure
on the NMRs,~\cite{Manipatruni2009} which comes into play because of the changes
in cavity lengths since the optomechanical couplings depend on the cavity 
lengths, i.e., $%
O_{mi}=\frac{\omega_{ai}}{L_{i}}\sqrt{\frac{\hbar}{m_{\text{eff},i}\omega_{mi}}}$
and  $%
O_{3i}=\frac{\omega_{a3}}{L_{3i}}\sqrt{\frac{\hbar}{m_{\text{eff},i}\omega_{mi}}}$
($i=1,2$), \cite{akram2015} where $L_{i}$ and $L_{3i}$ are the cavity lengths 
and $m_{\text{eff},i}$ is the effective mass of NMR. The scenario changes 
completely when the
optomechanical couplings satisfy $O_{m2} > O_{m1}$ as given in 
Figure 
	\ref{optomechanical coupling}b. When $O_{m1}=60$  MHz and $O_{m2}=1$  
	MHz, 
signal transmission from port 1 to 2 can be realized having no output at port  
1 around the origin. An amplification of output signal transmission can be seen 
which comes into play because of the constructive interference between 
different transmission paths, i.e., $a_1 \leftrightarrow b_1 \leftrightarrow 
a_3 \leftrightarrow b_2 \leftrightarrow a_2$  and $a_2 \leftrightarrow b_2 
\leftrightarrow a_3 \leftrightarrow b_1\leftrightarrow a_1$. \cite{du2020} The 
amplification of output field has been reported in cavity optomechanical 
systems multiple times using different cavity setups.
\cite{Liu2017,si2017,li2017,jiang2018,wen2019}
However, when the optomechanical couplings become equal/same, that is, $%
O_{m1} = O_{m2}$, the signal transmission behavior turns completely to
reciprocal as a result of symmetric radiation pressure on both the NMRs. For
example, when $O_{m1}= O_{m2}$, the signal is allowed
to pass through both output ports by the same amount at/around the origin on the
frequency axis (not shown here) which is reciprocal in nature.

In our proposed setup, the optomechanical couplings, i.e., 
$O_{m31}=O_{m32}=O_{m3}$ associated with cavity mode $a_3$ and NMRs are of 
great interest in realizing the conversion of reciprocal to non-reciprocal 
behavior of output fields. Non-reciprocity is valid if 
$O_{m3}$ is 
non-zero under the condition $O_{m1} \neq O_{m2}$. In the case, the coupling 
$O_{m3}$ goes down to zero, the non-reciprocity is lost, and the system is left 
with complete reciprocal signal transmission at output ports regardless of the 
values (higher or lower) assigned to couplings $O_{m1}$ and $O_{m2}$ as shown 
in Figure \ref{optomechanical coupling}c,d. Now, 
to see the output 
signal's nature turning from reciprocal to non-reciprocal, the system in this 
case rely on optomechanical coupling $O_{m3}$ value. Figure \ref{optomechanical 
	coupling}c,d show reciprocal signal transfer at both ports 2 and 1, 
respectively, having their maxima at the origin when $O_{m31}=O_{m32}= 
O_{m3}=0$ as expressed by black curves. The interference pattern is effectively 
modified as the coupling $O_{m3}$ step up to non-zero value, and thus the 
reciprocal nature of output signal at both ports is gradually transformed into 
non-reciprocal. The signal transfer T$_{1 \rightarrow 2}$  becomes dominantly 
amplified and its curve splits into two by further increasing the value of 
$O_{m3}$ thus gaining Fano-like steep peaks,  while the curve for 
T$_{2\rightarrow 1}$ also acquires a Fano-shaped profile 
keeping almost the 
same intensity. By further increasing the  value of $O_{m3}$ (8.5 MHz), the 
Fano-profiled asymmetric peaks of T$_{1 \rightarrow 2}$ become symmetric and 
skid away from the origin on frequency axis. On the other hand, the signal 
transfer T$_{2\rightarrow 1}$ stays on the origin and two 
small peaks 
smoothly converge to single peak. Beyond 8.5 MHz of $O_{m3}$ value, the 
transmission T$_{1 \rightarrow 2}$ curve profile at port 2 changes again from 
symmetric to asymmetric with its line-width being broadened thus showing 
effectively the output signal transfer away from origin. The output signal 
transmission T$_{2 \rightarrow 1}$ at port 1 owns the same curve pattern with 
single peak at constant intensity as shown in Figure 
	~\ref{optomechanical coupling}(d). From Figure 
	\ref{optomechanical coupling}c,d, it is obvious 
that by increasing the value of $O_{m3}$ the signal transfer at both ports 
gradually convert from reciprocal to non-reciprocal.
\begin{figure}[tbhp]
	\centering
	\includegraphics[width=0.40\linewidth]{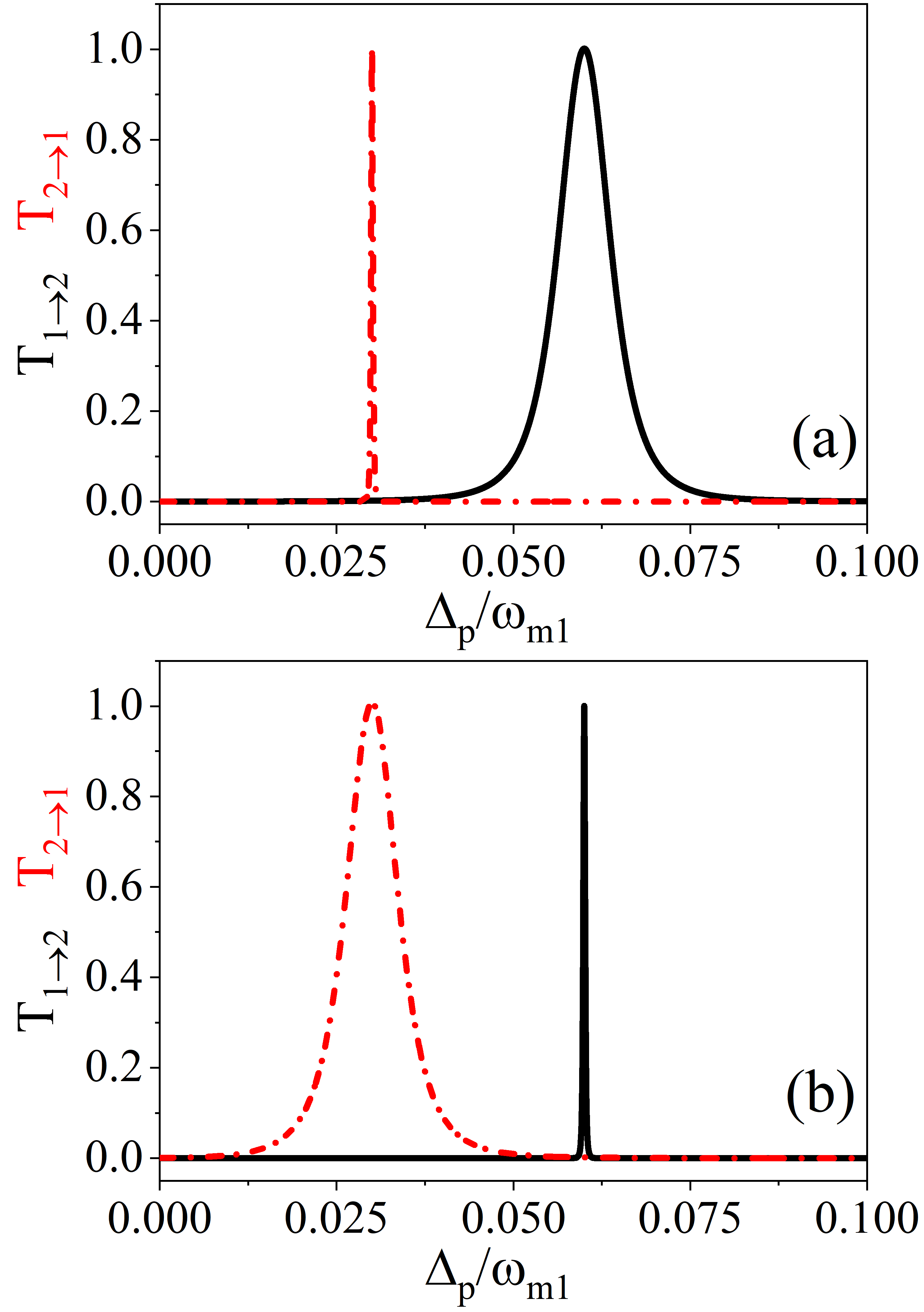}
	\caption{Probe transmission intensities T$_{2\rightarrow 1}$ (red dash-dot) 
		and T$_{1 \rightarrow 2}$ (black solid)  as a function of probe-drive 
		field 
		detuning
		under different values of cavity decay rates: (a) $\kappa_{1}/2\pi=83$ 
		MHz, $\kappa_{2}/2\pi=3$ MHz, $\kappa_{3}/2\pi=73$ MHz, and (b) 
		$\kappa_{1}/2\pi=3$ MHz, $\kappa_{2}/2\pi=83$ MHz, $\kappa_{3}/2\pi=73$ 
		MHz. The general parameters are
		given as, $\Delta_{1}=\Delta_{2}=\Delta_{3}=\protect\omega_{m1}$, 
		whereas other parameter values are same as in 
		Figure 2a}.
	\label{decay rate}
\end{figure}
Thus, the above discussion justifies the fact that signal
transmission at either port can be controlled flexibly from reciprocal to
non-reciprocal and vice versa by changing the 
optomechanical couplings.
\subsection{Effect of cavity decay rates on the signal flow}
Every COS inherits intrinsic photons dissipation to external bath (cavity decay 
rate) that depends on the
quality factor $Q$ of the end mirrors. Similarly, in our proposed cavity setup, 
the role of cavity decay rate $\kappa_{i}$ is inevitable and thus affects the
bidirectional signal transfer. As the transmission happens via left and right 
ports, we consider changes in the cavity decay rates associated with cavities 
$a_1$ and $a_2$ only. In \textbf{Figure \ref{decay 
rate}a}, when 
the cavity decay rates
have $\kappa_{1}>\kappa_{2}$, the system allows the output 
probe signal
from port 1 to port 2 with maximum value ( equal to 1) of T$_{1\rightarrow2}$ 
as shown by  black colored peak with large linewidth, but blocks
it in the opposite direction, i.e., from port 2 to port 1 with transmission 
value of T$_{2\rightarrow1}$ equal to zero. The above relation between two 
decay rates insinuate the razing of
photons in cavity $a_1$ as compared to cavity $a_2$ which eventually results in
suppressing of optical signals through cavity $a_1$ and thus to port 1. 
However, the
signal is transferred efficiently in reverse direction T$_{1 \rightarrow2}$. 
The larger amount of $\kappa_1$ is responsible for lowering the photon number 
and thus, the optomechanical coupling in cavity $a_1$ as compared to $\kappa_2$ 
value which results in comparatively larger optomechanical coupling in cavity 
$a_2$ and thus time reversal symmetry breaking happens that accounts for the 
non-reciprocal transmission. Due to the quantum interference mentioned in the 
above paragraphs, an ultra-thin peak can be achieved for  T$_{2\rightarrow1}$  
at a different frequency range where transmission in the opposite direction is 
zero thus showing the non-reciprocal behavior. Figure \ref{decay rate}b
reveals the case for signal transfer when $\kappa_{1}<\kappa_{2}$, so the
converse happens that provokes the signal transmission from port 2 to 1 T$%
_{2 \rightarrow 1}$ and suppresses it in the reverse direction, that is, T$_{1
	\rightarrow 2}$. Hence, the cavity with larger value of $\kappa$ blocks
signal transfer at its respective port coming from the 
other one, while cavity 
with lower decay rate supports signal transmission at the same port.  Thus, 
non-reciprocity can be observed by considering the cavity decay rate values 
where the signal transfer behavior can be manipulated.
\subsection{Effect of probe and drive phases on signal flow}
\begin{figure}[tbhp]
	\centering
	\includegraphics[width=0.40\linewidth]{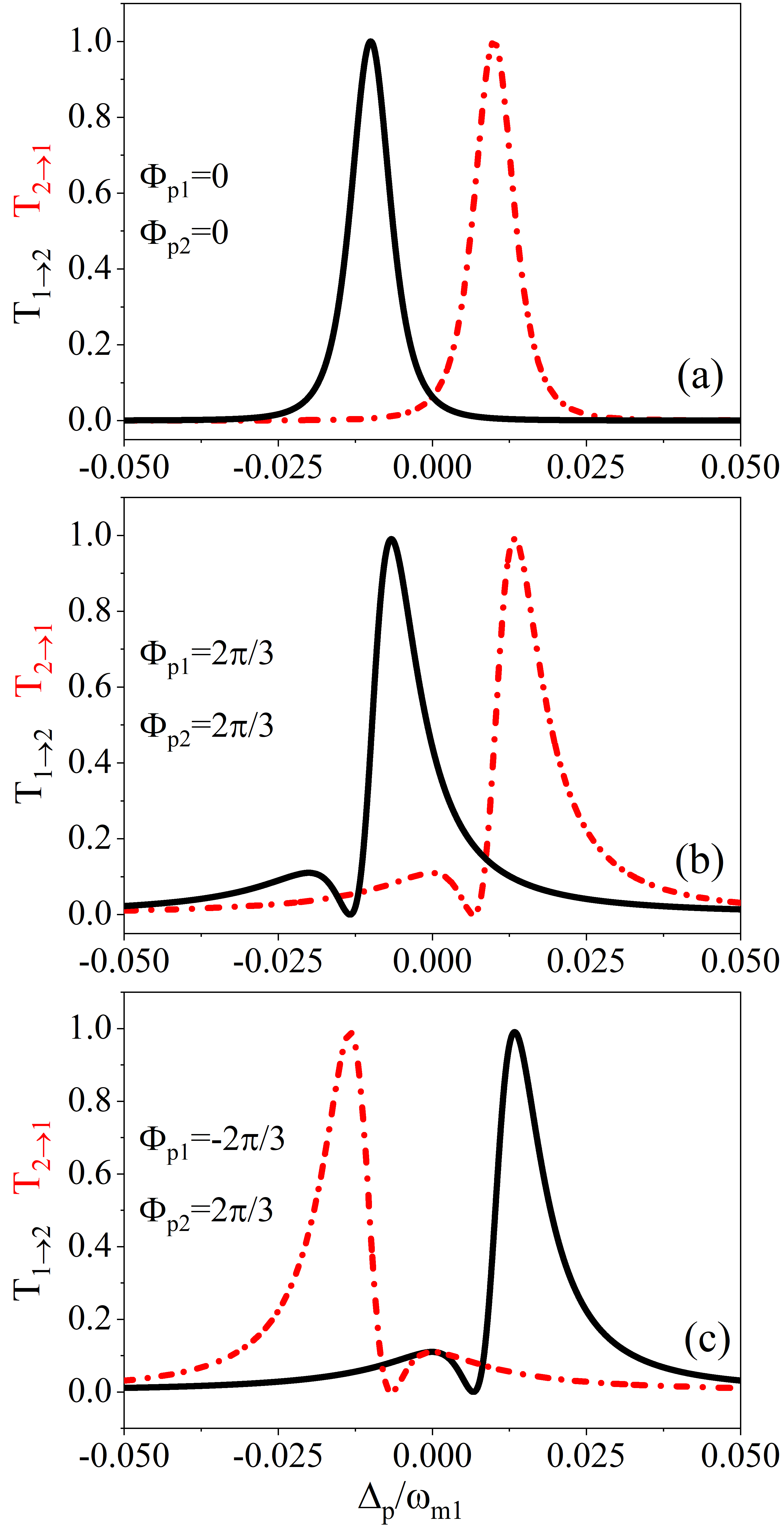}
	\caption{Dependence of transmission intensities T$_{2\rightarrow 1}$ (red 
		dash-dot) and T$_{1 \rightarrow 2}$ (black solid) on the probe-drive 
		field 
		detuning $\Delta_p$ when (a) probe phases $\Phi_{p1}=\Phi_{p2}=0$, (b) 
		$\Phi_{p1}=\Phi_{p2}=2\pi/3$, (c) $\Phi_{p1}=-2\pi/3,~ 
		\Phi_{p2}=2\pi/3$.
		The general parameters are given as 
		$\Delta_{1}=\Delta_{2}=\Delta_{3}=\protect\omega_{m1}$, and other 
		parameter values are
		same as in Figure 2(a).}
	\label{phase probe}
\end{figure}
\begin{figure*}[tbhp]
	\centering
	\includegraphics[width=0.90\linewidth]{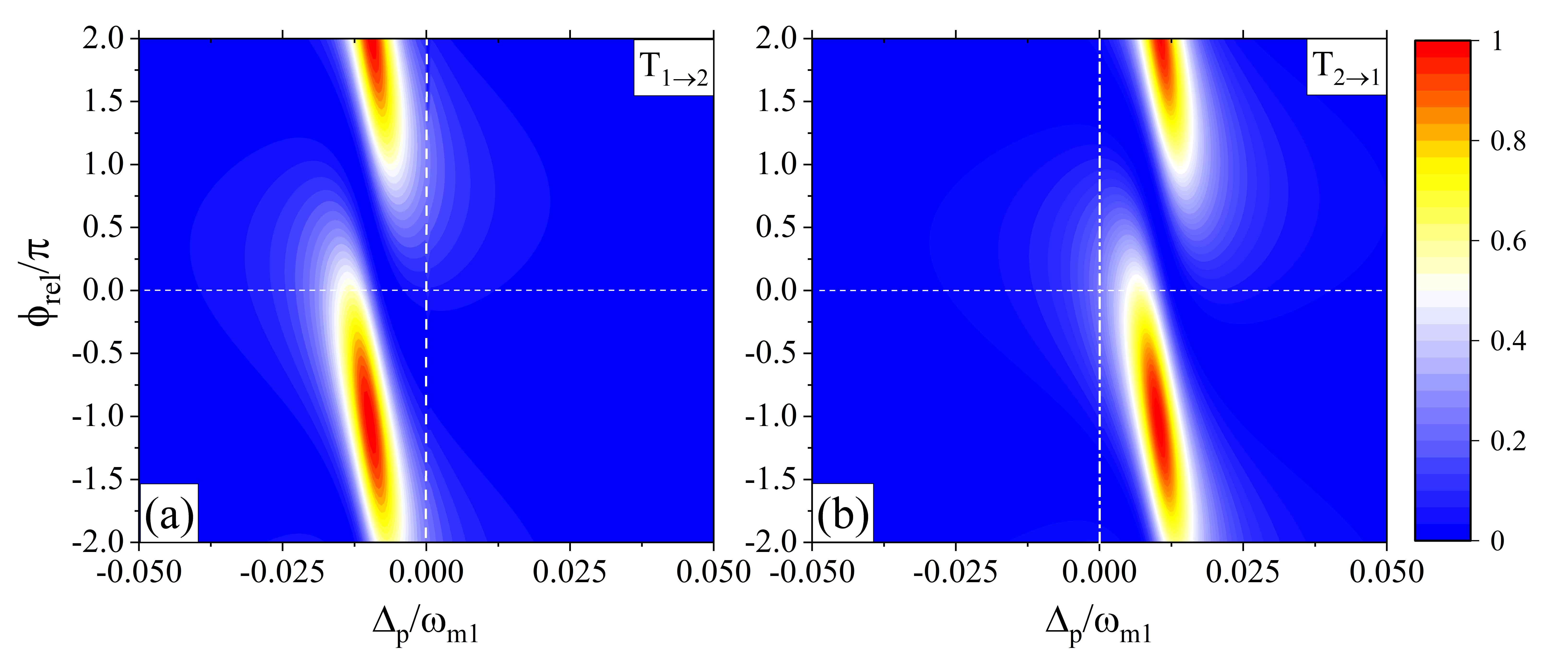}
	\caption{Density plot for the transmission intensities 
	(a) T$_{1
			\rightarrow 2}$ and (b) T$_{2
			\rightarrow 1}$ against the probe detuning $\Delta_{p}$ 
			and the relative phase $\phi_{\text{rel}}$. The general parameter 
			values are given 
			as 
		$\Delta_1=\Delta_2=\Delta_3=\omega_{m1}$. Other values are
		same as in Figure 2a}.
	\label{phase drive}
\end{figure*}
Generally, the phases of interacting fields play an important role in the 
interference phenomenon. Here we explain the significance of external probe and 
drive field phases that enables the cavity system to behave non-reciprocal for 
signal transmission at either port. These phase changes from both inputs are 
analogous to the
synthetic magnetic field that is responsible for breaking time reversal
symmetry and can be used as bidirectional non-reciprocal signal transport
device.~\cite{ruesink2018}
First, we suppose that the incoming external probe fields from either 
port 
have no phase change, that is, $\Phi_{p1}=\Phi_{p2}=0$. We report a complete 
non-reciprocal signal transfer via both ports with single transmission peak 
(black and red) at separate places on the frequency axis as shown in 
\textbf{Figure 
	\ref{phase probe}a}. Now, as the phases of probe fields are changed to 
$\Phi_{p1}=\Phi_{p2}=2\pi/3$, we still obtain non-reciprocal transmission 
curves with Fano-like profile  shown by Figure \ref{phase 
		probe}b. Since the 
signs of both the probe phases are positive, the curves attain the same profile 
at different positions on the probe detuning axis. As we change the sign of 
probe phase, i.e., $\Phi_{p1}=-2\pi/3$ and $\Phi_{p2}=2\pi/3$, the transmission 
T$_{1 \rightarrow 2}$ is shifted to the right side of origin having the same 
Fano-like shape as depicted by  Figure \ref{phase 
		probe}b. The output signal 
transfer T$_{2\rightarrow 1}$ shifts to the left of resonance position and 
reverses its direction due to the change in sign of probe phase as shown in 
Figure \ref{phase probe}c. The direction of 
non-reciprocal transmission 
completely reverses when the phases signs are changed, that  is,  
$\Phi_{p1}=2\pi/3$ and $\Phi_{p2}=-2\pi/3$ not presented here in figure. The 
same result given in Figure \ref{phase probe}c can be 
achieved for the phase 
changes equal to $\Phi_{p1}=\pi/3$ and $\Phi_{p2}=-\pi/3$.

Continuing the explanation of phases' role in controlling 
the behavior of 
signal transmission, we check the sensitivity 
of system to external fields phase on the 
non-reciprocal 
signal transfer. Since there are four electromagnetic fields (two drive and two 
probe fields) with corresponding phases $\Phi_{p1}$, $\Phi_{p2}$, 
$\Phi_{d1}$ and $\Phi_{d2}$ interacting with the cavity via two ports, it is 
vital to investigate the role of collective phase [also known as relative 
phase] in controlling the behavior of signal transferred at either port. The 
collective/relative phase can be expressed as $\phi_{\text{rel}}= 
(\Phi_{p1}+\Phi_{p2})-(\Phi_{d1}+\Phi_{d2})$, \cite{sahrai2004,ligang2008} and 
could be conveniently determined by recalculating Equation (\ref{delta-a1}) and 
(\ref{delta-a2}). Referring to \textbf{Figure \ref{phase drive}}, a density 
graph for 
the 
signal flow at the two ports has been plotted. Here we check the sensitivity of 
system to the relative phase $\phi_\text{rel}$ regarding the non-reciprocal 
signal 
transfer. Figure \ref{phase drive}a illustrates the transmission of signal from 
port 1 to 2 while Figure \ref{phase drive}b reveals the flow in reverse 
direction, that is, from port 2 to 1. By considering the relative phase, 
signal transmission T$_{1 \rightarrow 2}$ and T$_{2 \rightarrow 1}$ along the 
two opposite routes have a complete non-reciprocal nature which is depicted by 
the contours in Figure \ref{phase 
drive}a and \ref{phase drive}b, respectively. The contours demonstrate that 
when the relative phase is in the regions around 
$\phi_\text{rel}=-2 ~\text{to}~ 0$ and $\phi_\text{rel}=-1 ~\text{to beyond}~ 
2$ at the left frequency band of probe detuning, maximum signal transfer 
T$_{1 
\rightarrow 2}$ is achieved at its corresponding port 2, whereas the output 
signal T$_{2 \rightarrow 1}$ at port 1 is fully suppressed at that bandwidth of 
detuning frequency, thus, revealing its sensitivity upon the relative 
phase. The same explanation goes for the upper half contour where signal 
intensity lies. However, 
the signal transmission T$_{2 \rightarrow 1}$ at port 1 is maximum for the 
similar range of relative phase mentioned above on the right of resonance point 
on frequency 
axis, but there is no signal flowing in the opposite direction which again 
justifies the non-reciprocal behavior of output signal transfer. As mentioned 
in the discussion corresponding to Figure \ref{phase 
probe}, the non-reciprocal behavior of signal transmission taking place here is 
due to the quantum interference since a change in the relative phase  of the 
fields may alter the interference pattern, thus resulting in the 
non-reciprocity 
due to the time reversal symmetry breaking. The signal from both ports also 
overlap in a certain range of relative phase values, i.e., 
$\phi_{\text{rel}}=0.75 ~\text{to almost} ~1$ hence revealing the reciprocal 
nature 
of 
output signals.
The above explanation suggests that the output signal's nature strongly 
depends on the relative phase which can be switched from reciprocal to 
non-reciprocal and vice versa.
Thus, by changing the probe and/or drive field and relative phase we can 
flexibly control 
the bi-directional non-reciprocal nature of the output signal at either port.

From the above description, it is clear that our proposed 
COS setup is 
phase-sensitive and non-reciprocal signal transfer is possible by changing the 
phases of  external probe fields and drive fields.

\section{\textbf{Summary}}
\label{summary}
We have theoretically investigated the non-reciprocal 
behavior of the output 
probe fields through a
bidirectional multi-mode COS being driven by external classical fields. A 
perfect non-reciprocal
transmission of signal due to the breaking of time reversal symmetry has been 
revealed at
the effective cavity detunings $\Delta_{1}$, $\Delta_{2}$ close to
mechanical frequency, and a full duplex transmission is tunable by adjusting
the values of $\Delta_{1}$ and $\Delta_{2}$. By modifying and flexibly 
controlling the optomechanical
couplings, signal transfer has been blocked from passing via one port
(terminal) and passed on through other around resonance conditions. 
Additionally, the nature of output signal has been realized to transform from 
reciprocal to non-reciprocal by smoothly tuning the optomechanical coupling 
values. The non-reciprocal
signal transfer is also influenced by tuning the cavity decay rates, which are 
the
intrinsic parameters that cannot be omitted. Interestingly, the phase changes 
and the relative phase
associated with input probe and drive fields from either port have crucial
impact on the signal transport and the transmission from reciprocal to
non-reciprocal and vice versa. This scheme suggested may be practically 
feasible in laboratory since it is based on cavity setup that has already been 
realized in experiments and is rooted on realistic system parameters. We hope 
that the proposed theoretical model could be the right route
for experimentalists to explore a new and efficient way for manufacturing
non-reciprocal devices like routers, optical isolators, sensors, light
diodes, and full duplex signal transmitters and transducers.
\section*{Acknowledgements} 
This research is supported by the National Natural Science Foundation of
China (grant Nos. 11974309), Zhejiang Provincial Natural
Science Foundation of China under Grant No. LD18A040001, and the grant by
National Key Research and Development Program of China (No. 
2017YFA0304202).

\section*{Conflict of interest}
The authors declare no conflict of interest.

\section*{Data Availability Statement} 
The data that support the findings of this study are available from the 
corresponding author upon reasonable request.
\medskip

%

\begin{thebibliography}{99}
	\bibitem {sounas2017}  D. L. Sounas, A. Al$\grave{\rm{u}}$, \textit{Nat. 
	Photonics},  \textbf{2017} 
\textit{11} 774.

\bibitem {lorentz1896}  H. A. Lorentz, \textit{Amsterdammer Akademie der 
	Wetenschappen}, \textbf{1896},
\textit{4} 176; https://ci.nii.ac.jp/naid \\/10018471922/en/

\bibitem{levy2005} M. Levy,  \textit{J. Opt. Soc. Am. B} \textbf{2005}, 
\textit{22}, 254.

\bibitem{zaman2007} T. R. Zaman, X. Guo, R. J. Ram, \textit{Appl. Phys. 
	Lett.}  \textbf{2007}, \textit{90},
023514.

\bibitem{dotsch2005} H. Dotsch, N. Bahlmann, O. Zhuromskyy, M. Hammer,
L. Wilkens, R. Gerhardt, P. Hertel, A. F. Popkov, \textit{J. Opt.
	Soc. Am. B} \textbf{2005}, \textit{22} 240.

\bibitem {wang2009} Z. Wang, Y. Chong, J. D. Joannopoulos, M.   
Solja$\check{\rm{c}}$i$\grave{\rm{u}}$, 
\textit{Nature (London)}, \textbf{2009}, \textit{461}, 772.

\bibitem {khanikaev2010} A. B. Khanikaev, S. H. Mousavi, G. Shvets,
Y. S. Kivshar, \textit{Phys. Rev. Lett.} \textbf{2010}, \textit{105}, 
126804.

\bibitem {bi2011} L. Bi, J. Hu, P. Jiang, D. H. Kim, G. F. Dionne,
L. C. Kimerling, C. A.  Ross, \textit{Nat. Photonics}, \textbf{2011}, 
\textit{5} 758.

\bibitem{chin2013} J. Y. Chin, T. Steinle, T. Wehlus, D. Dregely, T. Weiss, 
V. I. Belotelov, B. Stritzker, H. Giessen, \textit{Nat. Commun.} 
\textbf{2013}, 
\textit{4}, 1599.

\bibitem {dai2012} D. Dai, J. Bauters, J. E. Bowers, \textit{Light Sci. 
	Appl.} \textbf{2012}, \textit{1}, e1.

\bibitem{estep2014} N. A. Estep, D. L. Sounas, J. Soric, A.  
Al$\grave{\rm{u}}$, \textit{Nat. Phys.} \textbf{2014}, \textit{10}, 923.

\bibitem{kerckhoff2015}  J. Kerckhoff, K. Lalumi$\grave{\rm{e}}$re,   
B. J. Chapman, A.  Blais, K. W.  Lehnert, \textit{Phys. Rev. Appl.} 
\textbf{2015}, \textit{4}, 
034002.

\bibitem{ranzani2014} L.  Ranzani, J.  Aumentado, \textit{New J. 
	Phys.} \textbf{2014}, \textit{16}, 103027.

\bibitem{white2015} T. C.  White, J. Y. Mutus, I.-C. Hoi, R. Barends, B. 
Campbell, Y. Chen, Z. Chen, B. Chiaro, A. Dunsworth, E. Jeffrey, J. Kelly, 
A. Megrant, C. Neill, P. J. J. O'Malley, P. Roushan, D. Sank, A. 
Vainsencher,  J. Wenner, S. Chaudhuri J. Gao, J. M. Martinis, \textit{Appl. 
	Phys. 
	Lett.} \textbf{2015}, \textit{106}, 242601.

\bibitem{macklin2015}  C. Macklin, K. O'Brien, D. Hover, M. E. Schwartz, V. 
Bolkhovsky, X. Zhang, W. D. Oliver, I. Siddiqi, \textit{Science}, 
\textbf{2015}, 
\textit{350}, 307.

\bibitem{hua2016} S. Hua, J. Wen, X. Jiang, Q. Hua, L. Jiang, M. Xiao,    
\textit{Nat. Commun.} \textbf{2016}, \textit{7}, 13657.


\bibitem{hafezi2012} M. Hafezi, P. Rabl, \textit{Opt. Express} 
\textbf{2012}, \textit{20}, 7672.

\bibitem{shen2016} Z. Shen, Y.-L. Zhang, Y. Chen, C.-L. Zou, Y.-F. Xiao, 
X.-B. Zou, F.-W. Sun, G.-C. Guo, C.-H. Dong, \textit{Nat. Photonics}, 
\textbf{2016}, \textit{10} 657.

\bibitem{meystre2013}  P. Meystre, \textit{Ann. Phys. (Berlin)} 
\textbf{2013},  \textit{525 No. 
	3},  215.

\bibitem{chen2020}  S.-S. Chen, S.-S. Meng, Hong Deng, G.-J. Yang, 
\textit{Ann. Phys. (Berlin)} \textbf{2020}, \textit{533}, 
2000343; https://doi.org/10.1002/andp.202000343

\bibitem{shen2018} Z. Shen, Y.-L. Zhang, Y. Chen, F.-W. Sun, X.-B. Zou, 
G.-C. Guo, C.-L. Zou, C.-H. Dong, \textit{Nat. Commun. 
} \textbf{2018}, \textit{9}, 1797.

\bibitem{habraken2012} S. J. M. Habraken, K. Stannigel, M. D.  Lukin, 
P.Zoller, 
\textit{New J. Phys.} \textbf{2012}, \textit{14}, 115004.

\bibitem{xu2015} X.-W. Xu, Y. Li,  \textit{Phys. Rev. A} \textbf{2015}, 
\textit{91}, 053854.

\bibitem{zhou2016} B.-Y. Zhou, G.-X.  Li,  \textit{Phys. Rev. A} 
\textbf{2016}, \textit{94}, 033809.

\bibitem{ondrej2019}O. Cernot$\grave{\rm{i}}$k, C. Genes, A. Dantan,  
\textit{Quantum Sci. Technol.} \textbf{2019}, \textit{4},  024002.

\bibitem{tan2017} H. Tan, W. Deng, Q. Wu, G. Li, \textit{Phys. Rev. A}  
\textbf{2017}, \textit{95}, 053842.

\bibitem{rahman2018} Ziauddin, M. U. Rahman, I. Ahmad, S. Qamar,  
\textit{Euro phys. Lett.} \textbf{2017}, \textit{120}, 24001; 
https://iopscience.iop.org/article/10.1209/0295-5075/120/24001.

\bibitem{ullah2019} M. Ullah, A. Abbas, J. Jing,  L.-G. Wang,  
\textit{Phys. Rev. A} \textbf{2019}, \textit{100}, 063833.

\bibitem{qu2013} K. Qu, G. S. Agarwal, \textit{Phys. Rev. A} \textbf{2013}, 
\textit{87}, 063813.

\bibitem{kipf2014} T. Kipf, G. S. Agarwal, \textit{Phys. Rev. A} 
\textbf{2014}, 
\textit{90}, 053808.

\bibitem{si2017} L.-G. Si, H. Xiong, M. S. Zubairy, Y. Wu,  \textit{Phys. 
	Rev. A} \textbf{2017}, \textit{95}, 033803.

\bibitem{ville2019} V. Bergholm, W. Wieczorek, T.-S.  
Herbr$\ddot{\rm{u}}$ggen, M. Keyl,  \textit{Quantum Sci. Technol.} 
\textbf{2019}, \textit{4},  034001.

\bibitem{nenad2017} N. Kralj, M. Rossi, S. Zippilli, R. Natali, A. 
Borrielli, G. Pandraud, E. Serra, G. D. Giuseppe, D. Vitali,  
\textit{Quantum Sci. 
	Technol.} \textbf{2017}, \textit{2},  034014.

\bibitem{peterson2017} G. A. Peterson, F. Lecocq, K. Cicak, R. W. Simmonds, 
J. Aumentado, J. D. Teufel, \textit{Phys. Rev. X} \textbf{2017}, 
\textit{7}, 031001.

\bibitem{bernier2017} N. R. Bernier, L. D. T$\acute{\rm{o}}$th,  
A. Koottandavida, M. A.  Ioannou, D. Malz, A. Nunnenkamp, A. K. Feofanov,
T. J. Kippenberg,  \textit{Nat. Commun.} \textbf{2017}, \textit{8}, 604.

\bibitem{barzanjeh2017} S. Barzanjeh, M. Wulf, M. Peruzzo, M. Kalaee, P. B. 
Dieterle, O. Painter, J. M. Fink, \textit{Nat. 
	Commun.} \textbf{2017}, \textit{8}, 953.

\bibitem{xu2016} X.-W. Xu, Y. Li, A.-X. Chen, Y.-X. Liu, \textit{Phys. 
	Rev. A} \textbf{2016}, \textit{93}, 023827.

\bibitem{tian2017} L. Tian, Z. Li,  \textit{Phys. Rev. A} \textbf{2017}, 
\textit{96},  
013808.

\bibitem{mercier2020} L. M. de L$\acute{\rm{e}}$pinay, C. F. 
Ockeloen-Korppi, 
D. Malz, M. A. Sillanp$\ddot{\rm{a}}$$\ddot{\rm{a}}$,  \textit{Phys. 
	Rev. Lett.} \textbf{2020}, \textit{125}, 023603.

\bibitem{liu2021} S. Liu, B. Liu, J. Wang, L. Zhao, W.-X. Yang,  
\textit{J. Appl. Phys.} \textbf{2021}, \textit{129}, 084504.

\bibitem{agarwal2010} G. S. Agarwal, S. Huang,  \textit{Phys. Rev. A} 
\textbf{2010},
\textit{81}, 041803(R).

\bibitem{akram2015} M. J. Akram, M. M. Khan, F. Saif,  \textit{Phys. Rev. 
	A} \textbf{2015}, \textit{92}, 023846.

\bibitem{he2021} Q. He, F. Badshah, L. Li, L. Wang, S.-L. Su, E. Liang,  
\textit{Ann. Phys.} (Berlin) \textbf{2021}, \textit{533(5)}  
2000612; https://doi.org/10.1002/andp.202000612.

\bibitem{vitali2007} D. Vitali, S. Gigan, A. Ferreira, H. R. 
B$\ddot{\rm{o}}$hm, P. Tombesi, A. Guerreiro, V. Vedral, A. Zeilinger, M. 
Aspelmeyer,  \textit{Phys. Rev. Lett.} \textbf{2007}, 
\textit{98}, 030405.
\bibitem{huang2009} S. Huang, G. S. Agarwal,  \textit{New J. Phys.} 
\textbf{2009}, \textit{11}, 103044.

\bibitem{zou2011} C.-L. Zou, X.-B. Zou, F.-W. Sun, Z.-F. Han, G.-C. Guo,  
\textit{Phys. Rev. A} \textbf{2011}, 
\textit{84}, 032317.

\bibitem{agarwal2014} G. S. Agarwal, S. Huang, \textit{New J. Phys.} 
\textbf{2014},
\textit{16}, 033023.

\bibitem{walls1994} D. F. Walls, G. J. Milburn, \textit{Quantum Optics},  
Springer-Verlag,
Berlin \textbf{1994}.

\bibitem{yan2014} X. B. Yan, C. L. Cui, K. H. Gu, X. D. Tian, C. B. Fu, J. 
H. Wu, \textit{Opt. Express} \textbf{2014}, \textit{22(5)}, 4886.

\bibitem{xu2018} X.-W. Xu, L. N. Song, Q. Zheng, Z. H. Wang, Y. Li, 
\textit{ Phys. Rev. A} \textbf{2018}, \textit{98}, 063845.

\bibitem{kharel2019} P. Kharel, G. I. Harris, E. A. Kittlaus, W. H. 
Renninger, N. T. Otterstrom, J. G. E. Harris, P. T. Rakich, \textit{Sci. 
	Adv.} \textbf{2019}, \textit{5}, eaav0582.

\bibitem{Manipatruni2009} S Manipatruni, J. T.  Robinson, M. Lipson,  
\textit{Phys. Rev. Lett.} \textbf{2009}, \textit{102}, 213903.

\bibitem{du2020} L. Du, Y.-T. Chen, Wu J.-H. Wu, Y. Li, \textit{Opt. 
	Express} \textbf{2020}, \textit{28}, 3647.

\bibitem{Liu2017} Y. L. Liu, R. Wu, J. Zhang, S. K. Ozdemir, L. Yang, F. 
Nori, Y. X. Liu, \textit{ Phys. Rev. A} \textbf{2017}, \textit{95(1)}, 
013843.

\bibitem{li2017} Y. Li, Y. Huang, X. Zhang, L. Tian, \textit{Opt. 
	Express} \textbf{2017}, \textit{25}, 18907.

\bibitem{jiang2018} C. Jiang, B. Ji, Y. Cui, F. Zuo, J. Shi, G. Chen,   
\textit{Opt. Express} \textbf{2018}, \textit{26}, 15255.

\bibitem{wen2019} W.-A. Li, G.-Y. Huang, Y. Chen, \textit{J. Opt. Soc. 
	Am. B} \textbf{2019}, \textit{36}, 306.

\bibitem{ruesink2018} F. Ruesink, J. P. Mathew, M. A.  Miri, A.  
Al$\grave{\rm{u}}$, E. Verhagen, \textit{Nat. Commun.} \textbf{2018}, 
\textit{9}, 1798.

\bibitem{sahrai2004}M. Sahrai, H. Tajalli, K. T. Kapale, M. S. Zubairy,
\textit{Phys. Rev. A} \textbf{2004}, \textit{70}, 023813.

\bibitem{ligang2008}L.-G. Wang, S. Qamar, S.-Y. Zhu, M. S. Zubairy, 
\textit{Phys. Rev. A} \textbf{2008}, \textit{77}, 033833.

\end{thebibliography}


\end{document}